\documentclass[superscriptaddress, twocolumn, nofootinbib]{revtex4-1}

\usepackage{amssymb,amsmath}
\usepackage{latexsym}
\usepackage{graphicx}
\usepackage{caption}
\usepackage{subcaption}
\usepackage{enumerate}
\usepackage{float}

\usepackage{color}
\usepackage{graphics}
\usepackage{epsfig}
\usepackage{bm}

\newcounter{mnotecount}[section]

\begin{document}

\newcommand{\dR}{\mathbb R}
\newcommand{\dC}{\mathbb C}
\newcommand{\dZ}{\mathbb Z}
\newcommand{\id}{\mathbb I}
\newtheorem{theorem}{Theorem}
\newcommand{\ud}{\mathrm{d}}
\newcommand{\mfn}{\mathfrak{n}}

\author{Przemys{\l}aw Ma{\l}kiewicz}
\email{Przemyslaw.Malkiewicz@ncbj.gov.pl}

\affiliation{National Centre for Nuclear Research,  00-681
Warszawa, Poland}
\author{Artur Miroszewski}
\email{Artur.Miroszewski@ncbj.gov.pl}

\affiliation{National Centre for Nuclear Research,  00-681
Warszawa, Poland}

\author{Herv\'{e} Bergeron}
\email{herve.bergeron@u-psud.fr} 
\affiliation{ISMO, UMR 8214 CNRS,
Univ Paris-Sud, France}

\date{\today}

\title[]{Quantum phase space trajectories with application to quantum cosmology}

\begin{abstract}
We develop an approach to quantum dynamics based on quantum phase space trajectories. The latter are built from a unitary irreducible representation of the symmetry group of the respective classical phase space. We use a quantum action functional to derive the basic equations. In principle, our formulation is equivalent to the Hilbert space formulation. However, the former allows for consistent truncations to reduced phase spaces in which approximate quantum dynamics can be derived. We believe that our approach can be very useful in the domain of quantum cosmology and therefore, we use the cosmological phase space example to establish the basic equations of this formalism. 
\end{abstract}

 \maketitle

\section{Introduction}

In this article we propose a certain trajectory approach to quantum mechanics and we develop it with an emphasis on its application to simple quantum cosmological systems. The essence of our approach is to reformulate the Schr\"odinger equation densely defined on vectors in the Hilbert space in terms of a set of Hamilton's equations in a phase space. The quantum states are represented by phase space points. The phase space variables encode both classical and nonclassical observables. By classical observables in this context we mean the expectation values of basic operators in a given quantum state. By nonclassical observables we mean all other phase space variables which encode such properties as the dispersions of basic observables and so on.

Our formalism builds on the notion of coherent states and in particular on the idea of semiclassical framework by J. Klauder \cite{klauderII,primer}. In fact, our formalism is a natural extension of the Klauder framework which is included as a special case. The basic tools behind our formalism are, first, the variational formulation of the quantum dynamics once a quantum Hamiltonian has been provided and, second, coherent states that are constructed with a unitary and irreducible representation of a minimal canonical group in the phase space of the respective classical model. The coherent states are used to incorporate the classical observables into our formalism in a natural (more precisely, covariant) way and, as it is in the original framework, the equations of motion for the classical observables include in general terms with nonvanishing $\hbar$. Nevertheless, our extension of the framework by nonclassical observables brings in some new advantages that are not present in the original framework.

In principle, our approach is based on an infinite-dimensional phase space representing the quantum Hilbert space. However, its main advantage is that it includes consistent truncations to finite-dimensional phase spaces. Specifically, the dimensionality can be as large as to reproduce the exact Schr\"odinger equation in the form of phase space trajectories in the case of infinitely many dimensions, or as small as to give the roughest quantum corrections to the classical equations of motion in the case of the classical phase space dimensionality. The latter case corresponds, in fact, to the Klauder framework. Our approach allows to control the level of detail and can be adjusted to any specific quantum system, in particular, it can be used to deal with those quantum dynamics which are too complex to be solved explicitly.

Since we are mainly interested in quantum cosmological systems, we will develop our ideas for the case of the classical phase space which is a half-plane rather than a plane. For this particular phase space, a well-suited minimal canonical group is the affine group. Nevertheless, we wish to emphasize that our approach can be easily developed for the case of Weyl-Heisenberg (W-H) and other groups. 

The great advantage of our approach is that it allows for addressing physical questions which are not possible in the simplest semiclassical framework. For example, as usual classical phase space is now extended by nonclassical observables (such as spreading variables which describe specific quantum effects) one can investigate the dynamical properties of nonclassical observables in pseudo-classical terms of energy transfers between classical and nonclassical observables. It is worth mentioning that  the appearance of such phenomenons could be misinterpreted as ``real'' energy transfers from/to hidden spatial dimensions in the universe, which are introduced by the brane-world theories and alike \cite{brane1,brane2}. In fact, a wave-function naturally spreads when a system leaves a classical-like regime and enters a more quantum one. The dynamical spreading is in particular expected for cosmological systems when they approach the big-bang singularity. As it was already shown in \cite{frw}, in this case a quantum repulsive potential may halt the contraction preventing the universe form collapsing into the singularity and make it bounce and re-expand. Despite the fact that the expectation values of the basic observables such as the volume or the Hubble rate evolve symmetrically on both sides of the bounce (see \cite{frw} or almost any other work on the semiclassical dynamics of bouncing Friedmann models), on the fully quantum level the bounce does not simply revert the evolution. Thus, the evolution of some of quantum features is expected to be asymmetric with respect to the bounce. The detailed behavior can be captured within our extended approach by inspecting the evolution of nonclassical observables.

Other associated questions that can be potentially examined within our approach include: How to specify the degree of ``classicality'' (i.e. a measure of classicality) of a quantum behavior or, put differently in the present context, what is the ``classical universe''? How this classical universe emerges from a quantum state? Had the universe been classical before the bounce? Note, however, that our approach concerns only the deterministic dynamical behavior of objects in quantum mechanics, which includes the wave-function itself and the expectation values of observables. Other deep questions involved in quantum mechanics such its interpretation or the measurement problem are not addressed within this framework.

To finish these introductory remarks, let us notice that our formulation offers an alternative way of looking at quantum dynamics. To some extend it could be also viewed as a kind of a hidden-variable theory. Our approach is based on the time-dependent variational principle and therefore, it bears some resemblance to the formalism developed in \cite{Kramer}. However, the latter lacks the physical interpretation that we obtain thanks to the use of coherent states.

The outline of the article is as follows. We begin by recalling in Sec II some basic properties of the expectation values in quantum mechanics, which sets a broad context for our framework. In Sec III we recall the semiclassical framework of Klauder. In Sec IV we discuss the variational formulation of quantum dynamics. In Sec V we develop our formalism. We apply it to two examples in Sec VI. In Sec VII we revisit the quantum flat Friedmann model with our approach. We conclude in Sec VIII. The appendices deal with some technicalities (self-adjointness and the numerical code) which have been omitted from the main text.

\section{States and expectation values in quantum mechanics}

Let us assume a quantum system described by a normalized state $| \psi \rangle$, or rather by the corresponding projector (ray) $P_\psi = | \psi \rangle \langle \psi |$ in a finite dimensional Hilbert space $\mathcal{H}$ of dimension $N$. $P_\psi$ belongs to the complex projective space $\mathbf{CP}^{N-1} \cong \mathbf{S}^{2N-1} / U(1)$ and depends on $2N - 2$ real parameters. For a quantum observable represented by a self-adjoint operator $\mathcal{O}$ on $\mathcal{H}$ the expectation value in a state $P_\psi$ is given by 
\begin{align}
\langle \mathcal{O} \rangle_\psi = {\rm Tr} \left( P_\psi \mathcal{O} \right) \in \mathbb{R}.\end{align}
The Lie algebra of self-adjoint operators on $\mathcal{H}$ is a real vector space of dimension $N^2$, or $N^2-1$ if we exclude the identity. Notice that $N^2-1 \ge 2N-2$ for $N \ge2$. Therefore, if we choose appropriately $2N-2$ independent observables $\{\mathcal{O}_i\}_{i=1}^{2N-2}$, the mapping 
\begin{align}
P_\psi \mapsto \vec{x}_\psi = \big\{\langle\mathcal{O}_1\rangle_\psi, \langle\mathcal{O}_2\rangle_\psi,\dots, \langle\mathcal{O}_{2N-2}\rangle_\psi\big\} \in \mathbb{R}^{2N-2},\end{align}
is locally invertible. Hence, the set of rays $P_\psi$ can be seen as a manifold locally parametrized by an array of expectation values $\vec{x} \in \mathbb{R}^{2N-2}$. This mapping gives a natural physical picture of a quantum state: {\bf a quantum state is a complete set of statistical properties specified by a family of expectation values}. The inverse mapping: $\vec{x} \mapsto P_{\vec{x}}$, allows to define any expectation value of any quantum observable $\mathcal{O}$ as a function 
\begin{align}\label{obs}
\vec{x} \mapsto f_{\mathcal{O}}(\vec{x}) := {\rm Tr} \left(P_{\vec{x}} \mathcal{O} \right).\end{align} Hence, the set of quantum expectation values looks like a set of classical observables defined on a classical phase space represented here by the set of $\vec{x}$. This picture is enhanced by the Ehrenfest theorem stipulating that expectation values have a deterministic behavior through equations similar to the Hamilton equations. Notice, however, that any function of $\vec{x}$ is not an expectation value of a quantum observable. This is different from the usual classical framework.

Notice that this picture obscures those quantum aspects of single systems that are addressed by the so-called measurement axioms of quantum mechanics. Nevertheless, the usual stochastic quantum reasoning remains in principle accessible since the quantum probabilities yielded by the Born rule,
\begin{align}
 | \langle \psi | \phi \rangle |^2 = {\rm Tr} (P_\psi P_\phi),\end{align}
are included in the framework through Eq. (\ref{obs}) for $P_\psi := P_{\vec{x}}$ and $P_\phi:=\mathcal{O}$.

The above picture is very attractive for establishing a bridge between classical and quantum calculations. Indeed, if we ignore the quantum stochastic origin of the picture, we recover a classical-like formalism. The presented construction is valid only for finite dimensional Hilbert spaces, though, the idea of using expectation values is obviously attractive for the infinite dimensional spaces as well. A desired extension can be established if one finds a way to truncate the principally infinite sequence of expectation values needed to specify a quantum state belonging to an infinite dimensional Hilbert space. Herein we propose a suitable framework. Let us emphasize that this framework bears little resemblance with the usual ``phase space formulation of quantum mechanics'' based on Wigner functions, Weyl-Wigner transformation, the star product, etc. 

\section{Coherent states and Klauder's framework}
\subsection*{Coherent states}
By coherent states (see e.g. \cite{gazeau})  we mean a continuous mapping from a set of labels, collectively denoted by $l$ and equipped with a measure $\ud l$, into unit vectors in Hilbert space,
\begin{align}
l\mapsto | l\rangle\in\mathcal{H},
\end{align}
such that it resolves the identity,
\begin{align}
\int\ud l~| l\rangle\langle l|=I_{\mathcal{H}}.
\end{align}
Hence, the coherent states $| l\rangle$ form an overcomplete basis in  $\mathcal{H}$. The above property was first used by  Klauder \cite{klauderI} in his definition of what he called an overcomplete family of states (OFS). They provide a bridge between the abstract quantum formalism and the continuous label-space,
\begin{align}
\mathcal{H}\ni|\psi\rangle\mapsto P_{\psi}(l):=|\langle l |\psi\rangle|^2,
\end{align}
where $P_{\psi}(l)$ is a normalized (with respect to $\ud l$) probability distribution on the space $l$ that can be further identified with some classical observables. Suppose that there exists a unitary irreducible representation on $\mathcal{H}$, $U(\mathcal{X})$, of a minimal group of canonical transformations in a phase space $\mathcal{X}$. Then, the mapping
\begin{align}
\mathcal{X}\ni \xi\mapsto |\xi\rangle:=U(\xi)|\psi_0\rangle\in\mathcal{H},
\end{align}
defines a family of coherent states whose labels describe the classical states of that system \cite{perelemov}. The fixed normalized vector $|\psi_0\rangle$ is called the fiducial vector.
 
\subsection*{Framework}
First, let us recall that the quantum dynamics can be obtained via the variation of the quantum action,
\begin{align}\label{QA}
S(\psi,\dot{\psi})=\int_{t_i}^{t_f}\ud t\langle\psi|i\partial_t-\hat{H}|\psi\rangle,
\end{align}
with respect to the normalized $|\psi\rangle\in\mathcal{H}$, where $\hat{H}$ is the quantum Hamiltonian that corresponds to a certain classical Hamiltonian, $H$. The stationary points of the action (\ref{QA}) are found to satisfy the Schr\"odinger equation,
\begin{align}
i\partial_t|\psi\rangle=\hat{H}|\psi\rangle .
\end{align}
The idea of the semiclassical framework based on the coherent states was introduced by Klauder in \cite{klauderII}. Initially, he applied it to the case of the phase space $\mathcal{X}=\mathbb{R}^2$ and the Weyl-Heisenberg coherent states. The W-H coherent states, $|x,p\rangle$, are defined as follows
\begin{align}
|x,p\rangle:=D(x,p)|\psi_0\rangle,
\end{align}
where the displacement operator $D(x,p)=e^{i(p\hat{Q}-x\hat{P})}$ satisfies
\begin{align}
D(x',p')\circ D(x,p)=e^{\frac{i}{2}(xp'-px')}D(x+x',p+p'),
\end{align}
and where $(\hat{Q},\hat{P})$ are the position and momentum operators \cite{compendium}. The fiducial vector $|\psi_0\rangle\in\mathcal{H}$ is fixed and its choice is almost arbitrary. The only condition that one imposes on the fiducial is the so called physical centering condition. It relates the classical observables and the expectation values of the respective operators by demanding
\begin{align}
\langle x,p|\hat{Q} |x,p\rangle=x,~~\langle x,p|\hat{P} |x,p\rangle=p,
\end{align}
and leads to the constraint,
\begin{align}
\langle \psi_0|\hat{Q} | \psi_0\rangle=0 =\langle  \psi_0|\hat{P} | \psi_0\rangle.
\end{align}

We find that
\begin{align}\label{dWH}
i\langle x,p|\ud |x,p\rangle= p\ud x,
\end{align}
where $\ud$ is the exterior derivative. This result can be guessed (up to the irrelevant total derivative) from the fact that the above one-form must be invariant with respect to the action of the W-H group. The same reasoning applies to all other canonical groups. Now the quantum action functional (\ref{QA}) can be evaluated on the family of coherent states,
\begin{align}\nonumber
S(x,p)&=\int_{t_i}^{t_f}\ud t\langle x,p|i\partial_t-\hat{H}|x,p\rangle\\
&=\int_{t_i}^{t_f}\ud t\left(\dot{x}p-H^s(x,p)\right),
\end{align}
where $H^s=\langle x,p|\hat{H}|x,p\rangle$. Thus, the variation of the quantum action with respect to the $(x,p)$-labelled coherent states yields the Hamilton equations for $x$ and $p$,
\begin{align}
\dot{x}=\frac{\partial H^s}{\partial p},~~\dot{p}=-\frac{\partial H^s}{\partial x}.
\end{align}
On the one hand, the above equation provides an approximation to the exact quantum motion via the coherent states,
\begin{align}
\mathbb{R}\ni t\mapsto |x(t),p(t)\rangle\in\mathcal{H},
\end{align}
and on the other hand, it establishes a very appealing interpretation of the {\it classical} observables and their dynamics within the more fundamental {\it quantum} framework. We notice that these equations, in general, differ from the classical equations as the {\it semiclassical} Hamiltonian $H^s$ may include $\hbar$-corrections,
\begin{align}
H_s=H+\mathcal{O}(\hbar).
\end{align}
Since in the real world $\hbar$ never vanishes, the possibility of modeling the dynamics of the classical observables with nonvanishing $\hbar$ could be very useful. Indeed, this possibility becomes particularly important for improving the dynamics of classically singular cosmological models as we show later.

\subsection*{Affine group}
As we are concerned with gravitational systems, we shall turn to the important example of the phase space that appears in cosmology, namely the half-plane $\mathcal{X}=\mathbb{R}_+\times\mathbb{R}$. The basic observables form a canonical pair,
\begin{align}
(q,p)\in\mathbb{R}_+\times\mathbb{R},
\end{align}
where $q$ is the volume of the universe and $p$ is a rate of its expansion. Clearly, the W-H group is not applicable to the present case as one of the canonical variables, $q$, is confined to the half-line. Instead, we shall employ the affine group \cite{affine,primer,affine2}, $A_{f}$, that is defined by the multiplication law,
\begin{align}
(q',p')\circ (q,p)=(q'q,\frac{p}{q'}+p'),
\end{align}
and preserves the symplectic structure of the half-plane phase space,
\begin{align}
(q',p')\circ [\ud q\wedge\ud p]=\ud (q'q)\wedge\ud(\frac{p}{q'}+p')=\ud q\ud p,
\end{align}
where $(q',p')$ is a fixed element of the affine group. There exists a unique (up to sign) unitary irreducible representation of $A_{f}$, which in $\mathcal{H}=L^2(\mathbb{R}_+,\ud x)$ takes the form
\begin{align}
U(q,p)\psi(x)=\frac{e^{ipx}}{\sqrt{q}}\psi\left(\frac{x}{q}\right).
\end{align}
Thus, we define the affine coherent states as
\begin{align}
|q,p\rangle:=U(q,p)|\psi_0\rangle,
\end{align}
where $\langle x|\psi_0\rangle\in L^2(\mathbb{R}_+,\ud x)$ is the fiducial vector that is subject to the constraint
\begin{align}
\int_{\mathbb{R}_+} |\psi_0|^2\frac{\ud x}{x}<\infty,
\end{align}
(which follows from the group integrability condition). To tighten the connection between quantum and classical observables we demand
\begin{align}\label{physicalcenteringAff}
\langle q,p|\hat{Q} |q,p\rangle=q,~~\langle q,p|\hat{P} |q,p\rangle=p,
\end{align}
where $\hat{Q}=x$ and $\hat{P}=\frac{1}{i}\partial_x$. This is equivalent to
\begin{align}
\langle \psi_0|\hat{Q} | \psi_0\rangle=1,~~\langle  \psi_0|\hat{P} | \psi_0\rangle=0.
\end{align}

One finds that
\begin{align}
i\langle q,p|\ud |q,p\rangle=-q\ud p+\langle\psi_0|\hat{D}|\psi_0\rangle\frac{\ud q}{q},
\end{align}
where $\hat{D}=\frac{1}{2i}\left(x\partial_x+\partial_x x\right)$ is the dilation operator and which confirms the general statement given below Eq. (\ref{dWH}).
Now, provided a quantum Hamiltonian $\hat{H}$, the quantum action functional evaluated on the affine coherent states reads
\begin{align}
S(q,p)=\int_{t_i}^{t_f}\ud t\left(\dot{q}p-H^s(q,p)\right),
\end{align}
where $H^s=\langle q,p|\hat{H}|q,p\rangle$. Hence, the variation of the quantum action with respect to the classical labels $(q,p)$ yields the Hamilton equations,
\begin{align}\label{affeom}
\dot{q}=\frac{\partial H^s}{\partial p},~~\dot{p}=-\frac{\partial H^s}{\partial q},
\end{align}
for the stationary trajectories.

\subsection*{Free particle dynamics on $q>0$}
\begin{figure}[t]
\centering
\includegraphics[width=0.3\textwidth]{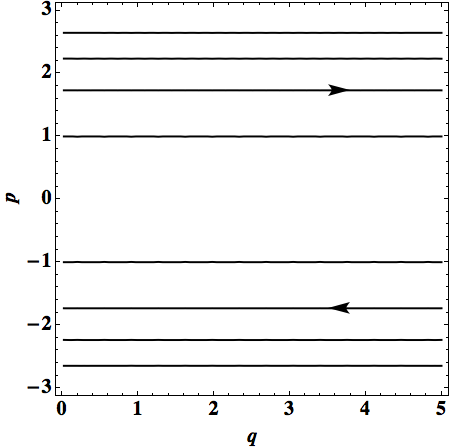}
\includegraphics[width=0.3\textwidth]{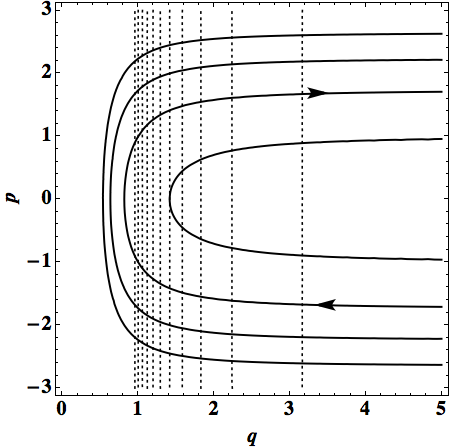}
\caption{Classical and semiclassical dynamics of a free particle on the half-line. As the semiclassical particle approaches the singular state $q=0$, it is repelled by the potential, which results in a bounce. We set $\hbar^2K=2$.}
\label{1}
\end{figure}

In this article we are going to study a quantum free motion of a particle on the half-line, $q>0$. It is a very important example as it formally describes the dynamics of the flat Friedmann universe with a perfect fluid-source \cite{frw}. The big-bang singularity is represented by the end-point, $q=0$. The variable $q$ describes the volume and the variable $p$ describes the expansion of the universe (see Sec VI for more details). Let the classical system be defined as follows,
\begin{align}
H=p^2,~~\omega=\ud q\ud p,~~(q,p)\in \mathbb{R}_+\times\mathbb{R},
\end{align}
and the quantum Hamiltonian read
\begin{align}\label{triangleoperator}
\hat{H}=-\bigtriangleup_x.
\end{align}
We discuss the technical issue of extending the above symmetric operator to a self-adjoint one in Appendix A. For a fiducial vector $\psi_0(x)\in L^2(\mathbb{R}_+,\ud x)$, we obtain
\begin{align}
S=\int_{t_i}^{t_f}\ud t\left(-\dot{q}p-H^s(q,p)\right),
\end{align}
where 
\begin{align}\label{simplebounce}
H^s = p^2+\hbar^2\frac{K}{q^2},
\end{align} 
where $K=\int_{\mathbb{R}_+}|\psi_0'|^2\ud x$. The respective Hamilton equations (\ref{affeom}) include an $\hbar^2$-correction that resolves the singularity at $q=0$. The particle is repelled away from the singularity by the quantum potential $\hbar^2\frac{K}{q^2}$ and this produces a bounce in its dynamics. See Fig. \ref{1}.

\section{Theory of the restricted quantum action}\label{restrictedaction}

The idea of this work is to extend the phase space description of quantum mechanics due to J. Klauder to a significantly broader framework. For this purpose it is useful to discuss the quantum action formulation of quantum dynamics in somewhat more detail.

\subsection*{Variation of the quantum action}
The quantum action is defined on trajectories in the Hilbert space and reads
\begin{align}\label{qafA}
S(\psi,\dot{\psi})=\int_{t_i}^{t_f}\ud t\langle\psi|i\partial_t+\triangle|\psi\rangle.
\end{align}
Its variation with respect to $\psi$ such that $\psi,\psi_{,x},\psi_{,xx}\in L^2(\mathbb{R}_+,\ud x)\cap C^{\infty}(\mathbb{R}_+)$ at each $t$, yields
\begin{align}\nonumber
\delta S=\int_{t_i}^{t_f}\ud t\int_{\mathbb{R}_+}(i\psi_{,t}\delta\bar{\psi}-i\bar{\psi}_{,t}\delta\psi)~\ud x\\\nonumber
+\int_{t_i}^{t_f}\ud t\int_{\mathbb{R}_+}(\psi_{,xx}\delta\bar{\psi}+\bar{\psi}_{,xx}\delta\psi)~\ud x\\ +\left[\int_{\mathbb{R}_+}i\bar{\psi}\delta\psi~\ud x\right]_{t_i}^{t_f}+\int_{t_i}^{t_f}\ud t\left[\bar{\psi}\delta\psi_{,x}-\bar{\psi}_{,x}\delta\psi\right]_0^{\infty}.
\end{align}
Provided that the variations vanish at the endpoints, $\delta\psi(t_i)=0=\delta\psi(t_f)$, the {\bf stationary points} of the quantum action $S(\psi,\dot{\psi})$ satisfy the Schr\"odinger equation,
\begin{align}
\left(i\partial_t+\triangle\right)\psi(x,t)=0.
\end{align}
We conclude that for each $\psi(x)$ there exists a unique stationary trajectory in the Hilbert space $\psi(x,t)$ such that $\psi(x,t_i)=\psi(x)$.

\subsection*{Variation of the reduced quantum action}

Now, suppose we confine the quantum action functional to trajectories in a subspace ${\Gamma}\subset L^2(\mathbb{R}_+,\ud x)$$\cap$ $C^{\infty}(\mathbb{R}_+)$ that is parametrized by real parameters. More precisely, we assume a differentiable map
\begin{align}
\mathbb{R}^{n}\ni\{\lambda_i\}\mapsto \psi_{\Gamma}\in \Gamma.
\end{align}
We will consider the reduced action
\begin{align}
S(\psi_{\Gamma},\dot{\psi}_{\Gamma})=\int_{t_i}^{t_f}\ud t\langle\psi_{\Gamma}|i\partial_t+\triangle|\psi_{\Gamma}\rangle.
\end{align}
Its variation yields
\begin{align}\nonumber
\delta S=\int_{t_i}^{t_f}\ud t\int_{\mathbb{R}_+}(i\psi_{{\Gamma},t}\delta\bar{\psi}_{\Gamma}-i\bar{\psi}_{{\Gamma},t}\delta\psi_{\Gamma})~\ud x\\+\int_{t_i}^{t_f}\ud t\int_{\mathbb{R}_+}(\psi_{{\Gamma},xx}\delta\bar{\psi}_{\Gamma}+\bar{\psi}_{{\Gamma},xx}\delta\psi_{\Gamma})~\ud x,
\end{align}
where $\delta\psi_{\Gamma}=\frac{\partial\psi_{\Gamma}}{\partial \lambda_i}\delta\lambda_i$ and $\delta\lambda_i(t_i)=0=\delta\lambda_i(t_f)$. The stationary trajectories $\psi_{\Gamma}(t)$ satisfy
\begin{align}\label{caeom}
\langle\delta\psi_{\Gamma}(t)|i\partial_t+\triangle|\psi_{\Gamma}(t)\rangle=0,
\end{align}
for any variations $\delta\lambda_i(t)$'s. In other words, {\bf the Schr\"odinger equation $\left(i\partial_t+\triangle\right)|\psi_{\Gamma}(t)\rangle=0$ holds only in the tangent space to $|\psi_{\Gamma}\rangle$}, namely
\begin{align}
T_{\vec{\lambda}}{\Gamma}=\textrm{span} \left(\frac{\partial|\psi_{\Gamma}\rangle}{\partial \lambda_i}\bigg|_{\vec{\lambda}}\right),
\end{align}
and, in general, $T_{\vec{\lambda}}{\Gamma}\neq \Gamma$. Given an orthonormal basis, $e_1,e_2,\dots, e_n$, in the tangent space $T_{\vec{\lambda}}{\Gamma}$, the equation of motion (\ref{caeom}) reads
\begin{align}
i\partial_t|\psi_{\Gamma}\rangle=\sum_i\langle e_i|-\triangle|\psi_{\Gamma}\rangle\cdot|e_i\rangle.
\end{align}
Suppose we gradually enlarge the subspace ${\Gamma}$ and its tangent space $T_{\vec{\lambda}}{\Gamma}$ by increasing the number of real parameters $\lambda_i$. Then, the orthonormal basis is enlarged accordingly $e_{n+1},e_{n+2},\dots$. Thus, for a fixed $|\psi_{\Gamma}\rangle$, its time derivative $i\partial_t|\psi_{\Gamma}\rangle$ becomes progressively a better and better approximation to the exact $-\triangle|\psi_{\Gamma}\rangle$ as the series
\begin{align}
\lim_{n\rightarrow\infty}\sum_{i=1}^n\langle e_i|\triangle|\psi_{\Gamma}\rangle\cdot|e_i\rangle= \triangle |\psi_{\Gamma}\rangle,
\end{align}
converges by the virtue of Parseval's identity. Notice that the convergence is defined for each point separately.

\subsection*{How to confine quantum motion?}
The Klauder semiclassical framework is based on a fixed family of coherent states. Each element of a given family satisfies the constraints,
\begin{align}
\langle q,p|\hat{Q}|q,p\rangle=q,~~\langle q,p|\hat{P}|q,p\rangle=p,
\end{align}
which tighten the relation between classical observables and their quantum counterparts. We may view the families of coherent states as sections of a certain fiber bundle \cite{bergeron}. Namely, the total space is the Hilbert space (or, its dense subspace), the base space is the space of all possible expectation values of the basic operators, and the fibers are made of state vectors that give equal expectation values,
\begin{align}
\pi:~\mathcal{H}\ni|\psi\rangle \mapsto( \langle \psi|\hat{Q}|\psi\rangle,\langle \psi|\hat{P}|\psi\rangle)\in \mathbb{R}_+\times\mathbb{R}.
\end{align}
The ``coherent'' sections are defined by fixing a fiducial vector $|\psi_0\rangle$ in the fibre $(1,0)$ and then by transporting it to all the other fibers via the unitary group action,
\begin{align}
|\psi_0\rangle\mapsto U(q,p)|\psi_0\rangle .
\end{align}
In other words, the orbits of the group define the ``coherent" sections. There are as many families of coherent states as fiducial vectors, $|\psi_0\rangle$, and the particular choice of the fiducial vector fixes purely quantum characteristics of the coherent states such as dispersions of the basic observables. They are nonclassical parameters that are completely fixed by the fiducial vector and are not allowed to evolve as they normally would do. Thus, the dynamical contribution from nonclassical observables is completely neglected in the Klauder framework and the only dynamical observables are the expectation values $(q,p)$ whose approximate dynamics could be for some purposes too rough. 

\begin{figure}[t]
\centering
\includegraphics[width=0.4\textwidth]{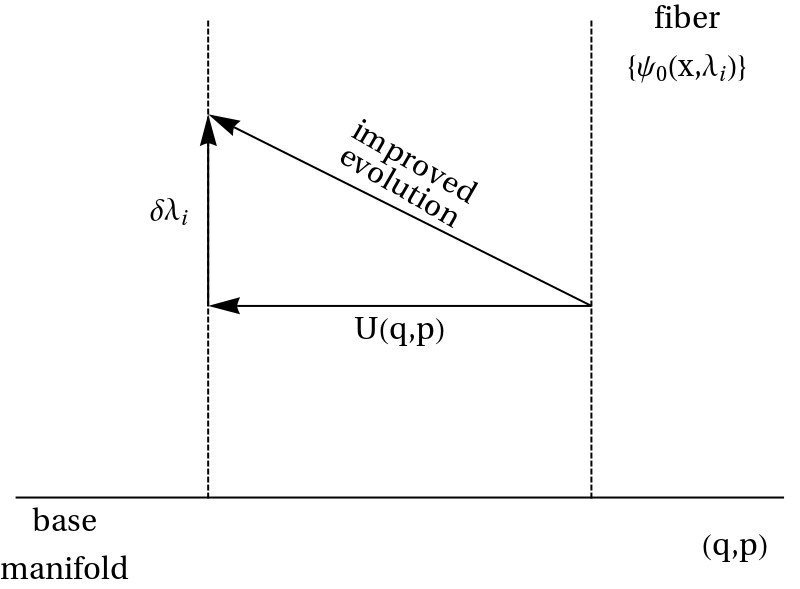}
\caption{We illustrate the quantum dynamics that takes place in the fiber bundle. The fibers consist of state vectors with the same expectation values, $q$ and $p$, of the basic operators, $\hat{Q}$ and $\hat{P}$, respectively. In the Klauder framework, it is the action of the affine group $U(q,p)$ that for given a fiducial vector $|\psi_0\rangle$ induces a section in the bundle, to which the quantum motion is confined. In our approach, we introduce extra parameters $\lambda_i$'s to parametrize the fiducial space. As a result, the quantum motion takes place both along the sections given by $U(q,p)|\psi_0\rangle$ and along the  fibers as the extra parameters can vary.}
\label{fiber}
\end{figure}

A way to improve this framework is to consider a fiducial space rather than a fiducial vector. It translates into confining the quantum motion to families of families of coherent states instead of a single family of coherent states. Such a framework allows the quantum motion to take place along the fibers of fixed expectation values of the basic observables. This idea is presented in Fig. \ref{fiber}. Such a framework would keep the connection between quantum states and classical observables while adding more dimensions to the phase space, which would describe purely quantum features of the quantum states. The number of the extra features would be controlled by the dimensionality of the fiducial space. Moreover, as we showed above, one expects that as the fiducial space is enlarged, the accuracy of this description is increased and it converges to a fully quantum mechanics expressed in terms of trajectories in a phase space of infinite dimension.

\section{Extension of Klauder's framework}

\subsection*{Quantum action}
We will now extend the Klauder semiclassical framework based on the affine coherent states. Instead of fixing a fiducial vector $|\psi_0\rangle$, we shall consider a fiducial space, $|\psi_0(\lambda_j)\rangle$, which contains the vectors labelled by $\lambda_j,~j=1,\dots$. Hence, we obtain a family of families of the affine coherent states,
\begin{align}
|q,p\rangle_{\vec{\lambda}}=U(q,p)|\psi_0(\lambda_j)\rangle,
\end{align}
which are labelled by $\lambda_j,~j=1,\dots$. The quantum action functional (\ref{qafA}) restricted to those families of the affine coherent states reads
\begin{align}
S=\int_{t_i}^{t_f}\ud t\left(-q\dot{p}+\frac{\dot{q}}{q}D-G^i\dot{\lambda}_i-(p^2+\frac{K}{q^2})\right),
\end{align}
where 
\begin{align}
G^i[\lambda_j]&=\langle\psi_0(\lambda_j)|\frac{1}{i}\partial_{\lambda_i}|\psi_0(\lambda_j)\rangle,\\
D[\lambda_j]&=\langle\psi_0(\lambda_j)|x\frac{1}{2i}\partial_x+\frac{1}{2i}\partial_xx|\psi_0(\lambda_j)\rangle,\\
K[\lambda_j]&=\langle\psi_0(\lambda_j)|-\triangle_x|\psi_0(\lambda_j)\rangle,
\end{align}
where we assume that $\psi_0(\lambda_j)\in L^2(\mathbb{R}_+,\ud x)\cap L^2(\mathbb{R}_+,\ud x/x)$ and that the conditions of normalization and for the expectation values for the basic observables hold,
\begin{align}\nonumber
&\langle\psi_0(\lambda_j)|\psi_0(\lambda_j)\rangle=1,~~\langle\psi_0(\lambda_j)|\hat{P}|\psi_0(\lambda_j)\rangle=0,\\ \label{physicalcenteringAff2}
&\langle\psi_0(\lambda_j)|\hat{Q}|\psi_0(\lambda_j)\rangle=1.
\end{align}
We assume the fiducial space to be linear and consist of the fiducial vectors of the form:
\begin{align}
|\psi_0(\lambda_j)\rangle=\sum\lambda_j|e_j\rangle,~~\lambda_j\in\mathbb{C},
\end{align}
such that $\langle e_j|e_i\rangle=N_{ji}$. Then,
\begin{align}
G^i[\lambda_j]&=\frac{1}{i}N_{ji}\bar{\lambda}_j,\\
D[\lambda_j]&=D_{ji}\bar{\lambda}_j\lambda_i,\\
K[\lambda_j]&=K_{ji}\bar{\lambda}_j\lambda_i,
\end{align}
where $N_{ji}$, $D_{ji}$ and $K_{ji}$ are hermitian. The quantum action reads now (after removing total time derivatives)
\begin{align}\label{actionfinal}
S=\int_{t_i}^{t_f}L\ud t,
\end{align}
where $L$ reads
\begin{align}
\left[\dot{q}\left(p+\frac{D_{ji}\bar{\lambda}_j\lambda_i}{q}\right)-\frac{1}{i}N_{ji}\bar{\lambda}_j\dot{\lambda}_i-\left(p^2+\frac{K_{ji}\bar{\lambda}_j\lambda_i}{q^2}\right)\right].
\end{align}
From the above action one derives the Hamiltonian formalism
\begin{align}
H&=p^2+\frac{K_{ji}\bar{\lambda}_j\lambda_i}{q^2},\\
\omega&=\ud q\ud\left(p+\frac{D_{ji}\bar{\lambda}_j\lambda_i}{q}\right)+\ud\lambda_i\ud(-\frac{1}{i}N_{ji}\bar{\lambda}_j)
\end{align}
with the quadratic constraints
\begin{align}\label{physicalcenteringAff3}
N_{ji}\bar{\lambda}_j\lambda_i=1,~~Q_{ji}\bar{\lambda}_j\lambda_i=1,~~P_{ji}\bar{\lambda}_j\lambda_i=0.
\end{align}
Note that the action (\ref{actionfinal}) yields the symplectic structure for both the classical and nonclassical observables. We follow the Dirac procedure \cite{dirac} and define the total Hamiltonian
\begin{align}
H_T=H+c_1N_{ji}\bar{\lambda}_j\lambda_i+c_2Q_{ji}\bar{\lambda}_j\lambda_i+c_3P_{ji}\bar{\lambda}_j\lambda_i,
\end{align}
where $c_i\in\mathbb{R}$ are to be determined with the use of the consistency conditions,
\begin{align}\nonumber
\partial_t(N_{ji}\bar{\lambda}_j\lambda_i)=\{N_{ji}\bar{\lambda}_j\lambda_i,H_T\}=0,\\
\partial_t(Q_{ji}\bar{\lambda}_j\lambda_i)=\{Q_{ji}\bar{\lambda}_j\lambda_i,H_T\}=0,\\\nonumber
\partial_t(P_{ji}\bar{\lambda}_j\lambda_i)=\{P_{ji}\bar{\lambda}_j\lambda_i,H_T\}=0.
\end{align}
\subsection*{Dynamics}
Since $N_{ji}$ is the identity operator and $D_{ji}$ is a hermitian operator, they can be simultaneously diagonalized. Suppose that they are diagonal, i.e. $N_{ji}= \delta_{ji}$, $D_{ji}= d_j\delta_{ji}$. Then,
\begin{align}
H_T&=p^2+\frac{{K}_{ji}\bar{\lambda}_j\lambda_i}{q^2}+c_1\delta_{ji}\bar{\lambda}_j\lambda_i+c_2{Q}_{ji}\bar{\lambda}_j\lambda_i+c_3{P}_{ji}\bar{\lambda}_j\lambda_j,\\
\omega&=\ud q\ud\left(p+\frac{d_j\delta_{ji}\bar{\lambda}_j\lambda_i}{q}\right)+i\ud\lambda_j\ud\bar{\lambda}_j.
\end{align}
We introduce $\gamma_j=\lambda_j e^{id_j\ln q}$ and find
\begin{align}
i\ud\gamma_j\ud\bar{\gamma}_j=i\ud\lambda_j\ud\bar{\lambda}_j-\ud q\ud\left(\frac{d_j\delta_{ji}\bar{\lambda}_j\lambda_i}{q}\right).
\end{align}
Thus, we may turn to a canonically equivalent formalism in which
\begin{align}
\omega=\ud q\ud p+i\ud\gamma_j\ud\bar{\gamma}_j,\\~~{K}_{ji}\bar{\lambda}_j\lambda_i=e^{i(d_j-d_i)\ln q}{K}_{ji}\bar{\gamma}_j\gamma_i=:k_{ij}\bar{\gamma}_j\gamma_i,\\
{Q}_{ji}\bar{\lambda}_j\lambda_i=e^{i(d_j-d_i)\ln q}{Q}_{ji}\bar{\gamma}_j\gamma_i=:q_{ij}\bar{\gamma}_j\gamma_i,\\~~{P}_{ji}\bar{\lambda}_j\lambda_i=e^{i(d_j-d_i)\ln q}{P}_{ji}\bar{\gamma}_j\gamma_i=:p_{ij}\bar{\gamma}_j\gamma_i.
\end{align}
and
\begin{align}
H_T=p^2+\frac{{k}_{ji}\bar{\gamma}_j\gamma_i}{q^2}+c_1\delta_{ji}\bar{\gamma}_j\gamma_i+c_2{q}_{ji}\bar{\gamma}_j\gamma_i+c_3{p}_{ji}\bar{\gamma}_j\gamma_i.
\end{align}
Let us define
\begin{align}
[MN]_{ji}\bar{\gamma}_j\gamma_i&:=\{M_{ji}\bar{\gamma}_j\gamma_i,N_{ji}\bar{\gamma}_j\gamma_i\}\\\nonumber
&=\frac{1}{i}(M_{jk}N_{ki}-N_{jk}M_{ki})\bar{\gamma}_j\gamma_i.
\end{align}
Now, the consistency relations yield
\begin{align}
c_1&= arbitrary~(phase~shift~generator),\\
c_2&=-\frac{(2p\frac{\partial p_{ji}}{\partial q}+\frac{1}{q^2}[pk]_{ji})\bar{\gamma}_j\gamma_i}{[pq]_{ji}\bar{\gamma}_j\gamma_i},\\
c_3&=-\frac{(2p\frac{\partial q_{ji}}{\partial q}+\frac{1}{q^2}[qk]_{ji})\bar{\gamma}_j\gamma_i}{[qp]_{ji}\bar{\gamma}_j\gamma_i}.
\end{align}
It follows that the normalization condition is a first-class constraint that generates a pure gauge transformation (an overall phase-shift) and thus, the coefficient $c_1$ is arbitrary. On the other hand, the physical centering conditions are second-class and the vaules of the coefficients $c_2$ and $c_3$ are determined. The equations of motion take the form
\begin{align}\label{eom}
\dot{q}&=2p,\\\label{eom2}
\dot{p}&=2\frac{{k}_{ji}\bar{\gamma}_j\gamma_i}{q^3}-\frac{{k}_{ji,q}\bar{\gamma}_j\gamma_i}{q^2}-c_2{q}_{ji,q}\bar{\gamma}_j\gamma_i-c_3{p}_{ji,q}\bar{\gamma}_j\gamma_i,\\\label{eom3}
\dot{\gamma}_{j}&=-i\frac{{k}_{ji}\gamma_i}{q^2}-ic_1\delta_{ji}\gamma_i-ic_2{q}_{ji}\gamma_i-ic_3{p}_{ji}\gamma_i.
\end{align}
\subsection*{A basis for the fiducial space}
In what follows we propose a set of orthonormal vectors $|e_i\rangle,~~i=0,1,2\dots$ that diagonalize the dilation operator $D_{ij}$. Observe the following unitary transformation:
\begin{align}
L^2(\mathbb{R}_+,\ud x)\ni \psi(x)\mapsto \phi(y)=e^{y/2}\psi(e^y) \in L^2(\mathbb{R},\ud y)
\end{align}
It transforms the dilation, position and momentum operator as follows
\begin{align}
\hat{D}=x\frac{1}{2i}\partial_x+\frac{1}{2i}\partial_xx &\mapsto \frac{1}{i}\partial_y,\\
\hat{Q}=x &\mapsto e^y, \\
\hat{P}=\frac{1}{i}\partial_x &\mapsto \frac{1}{i}e^{-y/2}\partial_y e^{-y/2}
\end{align}
Notice that the dilation operator is the momentum operator on $y\in\mathbb{R}$. Let us take the harmonic oscillator eigenvectors:
\begin{align}\label{basis}
\langle y|\psi_n\rangle=\frac{1}{\sqrt{2^n n!}}\frac{e^{-\frac{y^2}{2}}}{\sqrt[4]{\pi}}H_n(y)
\end{align}
where $H_n$ are the Hermite polynomials. If we restrict the considerations to the even eigenvectors, i.e. 
\begin{align}\label{bvs}
|e_n\rangle=|\psi_{2n}\rangle,
\end{align}
we obtain 
\begin{align}
N_{ij}=\langle e_{i}|e_{j}\rangle=\delta_{ij},~~D_{ij}=\langle e_{i}|\frac{1}{i}\partial_y|e_{j}\rangle=0.
\end{align}
In this case the dynamical analysis becomes very simple. Indeed, the equations of motion (\ref{eom},\ref{eom2},\ref{eom3}) become
\begin{align}
\dot{q}&=2p,\\
\dot{p}&=2\frac{K_{ji}\bar{\lambda}_j\lambda_i}{q^3},\\
\dot{\gamma}_{j}&=-i\frac{{K}_{ji}\lambda_i}{q^2}-ic_1\delta_{ji}\lambda_i-ic_2{Q}_{ji}\lambda_i-ic_3{P}_{ji}\lambda_i,
\end{align} 
where $c_1$ is arbitrary and
\begin{align}
c_2&=-\frac{[PK]_{ji}\bar{\lambda}_j\lambda_i}{q^2[PQ]_{ji}\bar{\lambda}_j\lambda_i},~~
c_3&= -\frac{[QK]_{ji}\bar{\lambda}_j\lambda_i}{q^2[QP]_{ji}\bar{\lambda}_j\lambda_i}.
\end{align}

\section{Numerical examples}

In what follows we consider two simple examples. In the first example, we set the fiducial space to be two-dimensional,
\begin{align}
|\psi_0\rangle=\lambda_1|e_1\rangle+\lambda_2|e_2\rangle,
\end{align}
where the vectors $|e_i\rangle$ are defined by Eq. (\ref{bvs}) and $\lambda_i\in\mathbb{C}$. We find that the absolute values $|\lambda_i|$ are constant in time while the respective phases are dynamical. The classical observables $q$ and $p$ undergo a simple bounce as in the case of Eq. (\ref{simplebounce}) to which solutions are presented in Fig. \ref{1}. This result is not surprising as there is, in fact, no extra degree of freedom. The counting of the extra degrees of freedom gives: 4 (two complex parameters) - 2 (two second-class constraints from the physical centering) - 2 (a first-class constraint and the respective gauge transformation from the normalization condition) = 0.

\begin{figure}
\centering
\includegraphics[width=0.36\textwidth]{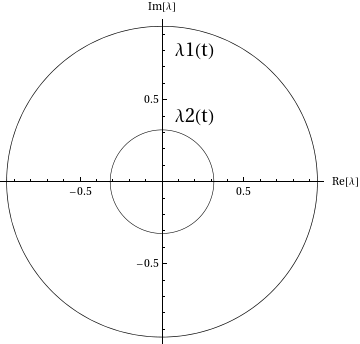}
\includegraphics[width=0.36\textwidth]{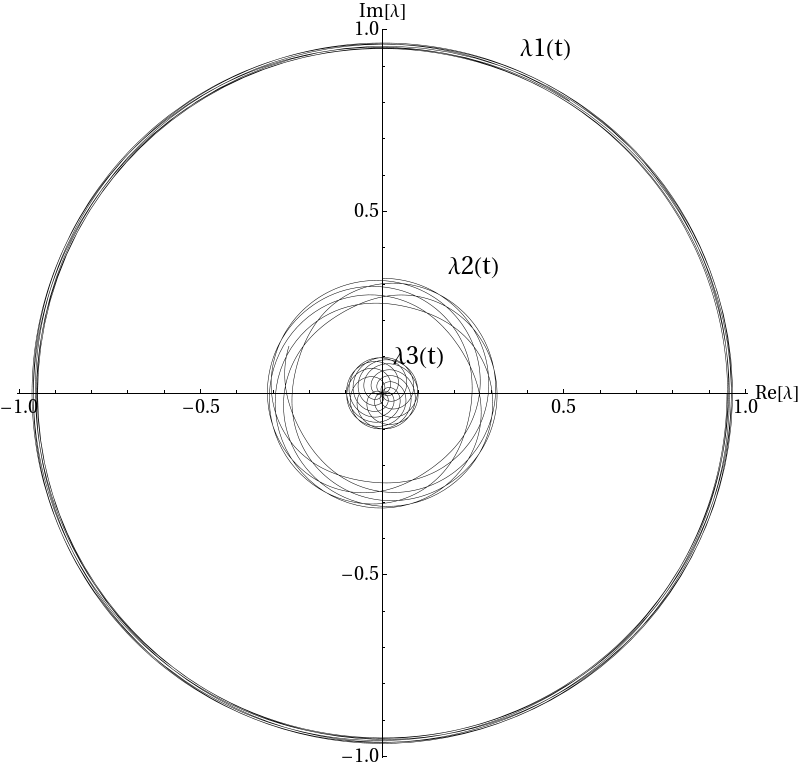}
\includegraphics[width=0.37\textwidth]{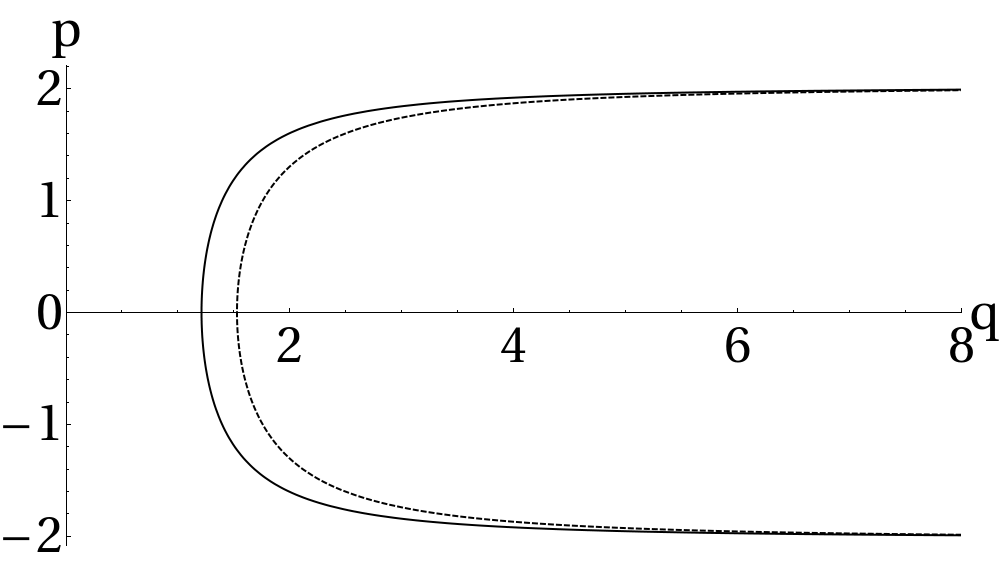}
\caption{We compare the cases of two and three extra complex parameters, $\lambda_1, \lambda_2$ and $\lambda_1, \lambda_2, \lambda_3$, respectively. The two upper plots show the dynamics of the extra parameters. For the two-parameter case, the extra parameters can only rotate in the complex plane. For the three-parameter case, the extra parameters exhibit very rich dynamics with both rotation and contraction/expansion. The latter proves that the evolution occurs across a set of families of coherent states. The bottom plot shows the dynamics of the classical observables $q$ and $p$ and despite the fact that the initial conditions for these observables are the same, the two-parameter (dashed) trajectory gives a bounce at smaller values of $q$ than the three-parameter (solid) one. As the initial condition we set $\lambda_1(0)=\sqrt{\frac{9}{10}}$, $\lambda_2(0)=-\sqrt{\frac{1}{10}}$, $\lambda_3(0)=0$, $q(0)=10$ and $p(0)=-2$.}
\label{Ex2}
\end{figure}

In the second example, we set the fiducial space to be three-dimensional,
\begin{align}
|\psi_0\rangle=\lambda_1|e_1\rangle+\lambda_2|e_2\rangle+\lambda_3|e_3\rangle.
\end{align}
In this case neither the absolute values $|\lambda_i|$ nor the respective phases are preserved during the evolution. In Fig. \ref{Ex2} we compare the dynamics of the classical observables and of the extra parameters between the two- and three-parameter cases.

\section{Quantum dynamics of the Friedmann universe}

\begin{figure}[h]
\centering
\includegraphics[width=0.45\textwidth]{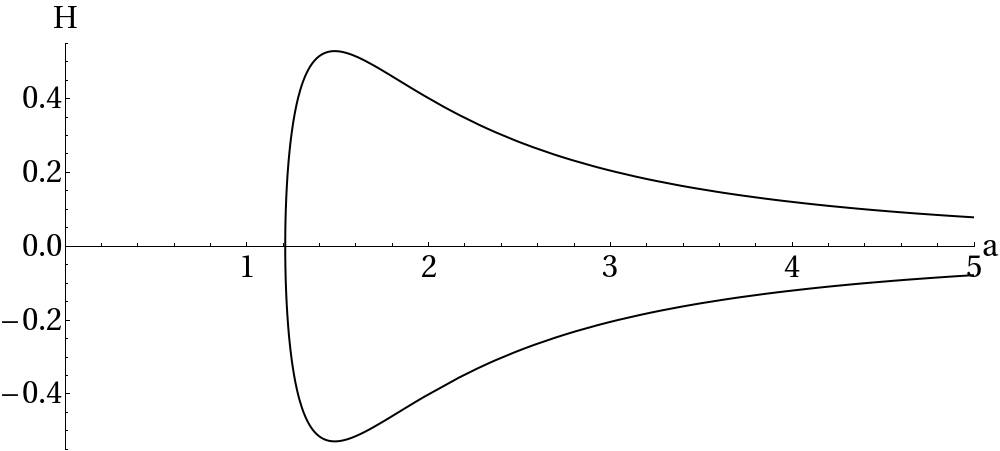}
\includegraphics[width=0.45\textwidth]{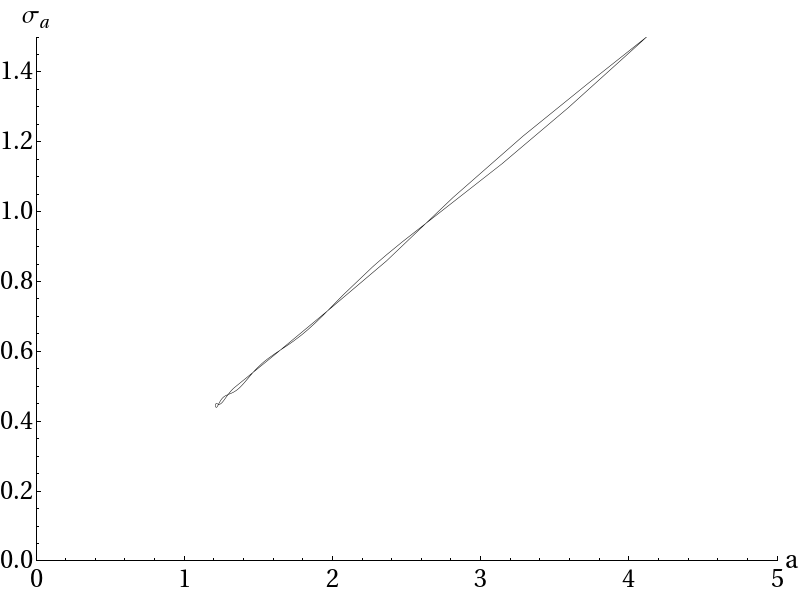}
\includegraphics[width=0.45\textwidth]{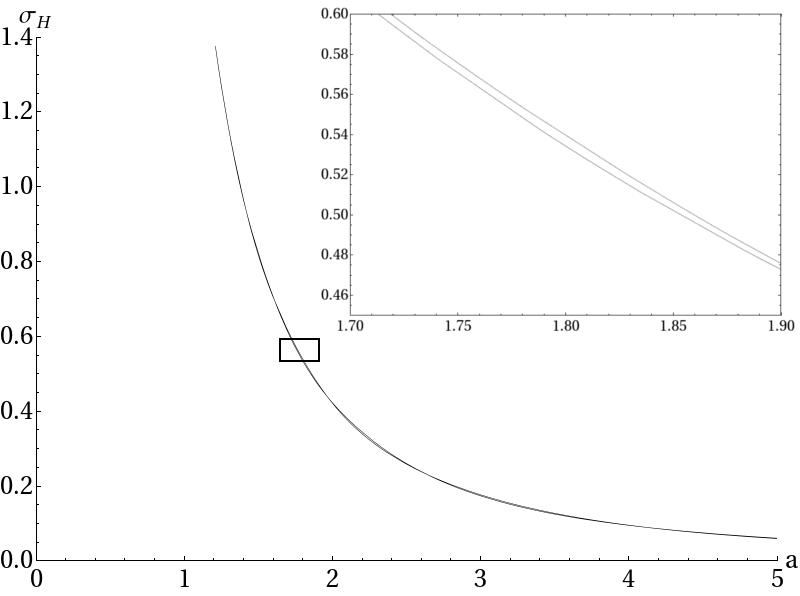}
\caption{The top plot shows the bouncing evolution of the Friedmann universe in the half-plane $(a,H)$. The two lower plots show the evolution of the dispersions $\sigma_a$ and $\sigma_H$ of the scale factor and the Hubble rate, respectively. We see the first evidence that the dynamics is not symmetric in time around the bounce. As the initial condition we set the initial data from the three-parameter case of Sec V, i.e $\lambda_1(0)=\sqrt{\frac{9}{10}}$, $\lambda_2(0)=-\sqrt{\frac{1}{10}}$ and $\lambda_3(0)=0$.}
\label{Ex3}
\end{figure}

Let us see how one can apply the formalism developed above to a quantum cosmological model, namely the quantum radiation-filled flat Friedmann universe with a bounce. For more details on the framework we refer to \cite{frw}. The metric of the classical model reads:
\begin{align}
\ud s^2=-N^2\ud t^2+q^2(\ud \vec{x})^2,
\end{align}
where $N$ is a nonvanishing and otherwise arbitrary lapse function. The Hamiltonian constraint reads
\begin{align}
\mathcal{C}=Nq^{-1}\left(-p^2+p_T\right),
\end{align}
where $T$ and $p_T$ are canonical variables that describe the radiation and
\begin{align}
q=a,~~p=a^2H,
\end{align}
are canonical variables that describe the geometry, the scale factor $a$ and the Hubble rate $H$ times the scale factor squared, respectively. We solve the Hamiltonian constraint with respect to $p_T$, set the lapse function $N=q$ and employ the variable $T$ as the internal clock. Then, the reduced phase space is given just by the canonical pair $(q,p)\in\mathbb{R}_+\times\mathbb{R}$ and the physical Hamiltonian reads
\begin{align}
H=p^2.
\end{align}
The above Hamiltonian can be promoted to the quantum Hamiltonian of Eq. (\ref{triangleoperator}). Then, we can use our approach to determine the quantum dynamics of the Friedmann universe in terms of a trajectory. In Fig. \ref{Ex3} we will plot the dynamics of the classical variables $a$ and $H$ and their dispersions. Note the following relations,
\begin{align}
\sigma_q&=\sqrt{\langle q,p|\hat{Q}^2|q,p\rangle-\langle q,p|\hat{Q}|q,p\rangle^2},\\
\sigma_p&=\sqrt{\langle q,p|\hat{P}^2|q,p\rangle-\langle q,p|\hat{P}| q,p\rangle^2},\\
\sigma_a&=\sigma_q,~~\sigma_H=\sqrt{4\frac{p^2}{q^6}\sigma_q^2+q^{-4}\sigma_p^2}.
\end{align}
In this section we present just one example of the extended phase space formulation of a quantum cosmological model. In our future work \cite{new} we investigate the quantum Friedmann model much more thoroughly.

\section{Conclusion}

In this article we present a quantum phase space trajectory approach to quantum dynamics. We start from the semiclassical framework introduced by J. Klauder many years ago and we extend it by inclusion of nonclassical observables that are equipped with a symplectic form. The obtained infinite-dimensional phase space trajectories are, in principle, equivalent to the exact solutions of the Schr\"odinger equation, though it is the possibility for consistent truncations to finite phase spaces that makes our approach attractive. We show that the respective Hamilton equations are not too complicated and can be successfully used for numerically integrating the dynamics.

Our trajectory approach is a tool that opens new possibilities in the studies of quantum cosmological systems. In the present article we test our approach with two simple examples. We postpone a detailed study of cosmological systems to our next papers. We believe that our approach can be helpful in establishing a definition of the ``degree of classicality'' of cosmological systems. If the universe is quantum by nature, it is never really classical or, put differently, quantum mechanics cannot ``disappear". Therefore, ``classicality'' must correspond to a special quantum behavior or, more precisely, to a special behavior in a particular picture of quantum dynamics. Given such a definition, we may be able to ``explain'', or ``recover'', the supposed classicality of the present universe and probe the effects of the lack of classicality on the past of the universe. We may learn if the universe can move back and forth between the classical and quantum phases. Finally, we could verify whether the universe could had been classical before the bounce. We investigate these and other related issues in the forthcoming paper \cite{new}.

Since the main purpose of developing this framework was to study quantum cosmological systems, we are led to ask to what extent a framework based on expectation values can reasonably describe a single system, namely the universe. There are two possible attitudes. The first attitude is to focus on the mathematical structure. Since our framework includes the complete time behavior of the wave-function, it is physically equivalent to the Schr\"{o}dinger equation. If our main purpose is to construct a picture of quantum dynamics that allows for a direct comparison with the corresponding classical equations of motion, we simply state that our framework is a very good candidate. The second attitude starts with the observation that the question of the interpretation of our formulation has been ignored. However, a similar interpretational issue arises in statistical physics where it is addressed with the so-called thermodynamical limit. Namely, statistical physics is designed to describe \emph{ensembles} of systems in terms of probabilities. However, the expectation values obtained from this theory are able to describe individual large systems: this is the thermodynamical limit. We can make an analogy and view a homogeneous cosmological system as made of an infinite number of ``copies'' of the same system localized at different points of space and therefore, describable as a large system made of ``small identical systems''.  Provided that the property valid in the framework of statistical physics can somehow be applied to the quantum cosmological context, the set of quantum expectation values in the cosmological framework becomes a relevant  description of the universe.

As a final remark, let us make a brief comparison of our approach to the Bohm-de Broglie (BdB) approach used in quantum cosmology \cite{pinto-neto}. In the BdB formulation, a given solution to the Schr\"{o}dinger equation plays the role of the so-called pilot-wave which is a source of an extra quantum term in the classical equations of motion. The latter determine a complete set (i.e., for arbitrary initial data) of quantum trajectories in the classical phase space. Whereas in our approach, a given solution to the Schr\"{o}dinger equation is represented by a unique trajectory of quantum expectation values in an infinite dimensional phase space that includes both classical and nonclassical variables. Thus, in our approach the system follows a unique and predictable, though, higher dimensional trajectory.

\section*{Acknowledgments} 
A. M. was supported by Narodowe Centrum Nauki with Decision No. DEC-2012/06/A/ST2/00395.

\section*{Appendix A}
The symmetric operator of Eq. (\ref{triangleoperator}),
\begin{align}
-\frac{\partial^2}{\partial x^2},~~D\left(-\frac{\partial^2}{\partial x^2}\right)=C_c^{\infty}(\mathbb{R}_+),
\end{align}
defined on smooth functions with a compact support is symmetric. There exist infinitely many self-adjoint extensions of the operator, which can be obtained by extending the domain to $C^{\infty}(\mathbb{R}_+)\cap L^2(\mathbb{R}_+,\ud x)$ with the boundary condition,
\begin{align}
\psi'(0)+\mu\psi(0)=0,
\end{align}
where $\mu\in\mathbb{R}\cup\{\infty\}$ labels the extensions \cite{reedsimon}. In the article, we impose the Dirichlet condition, $\psi(0)=0$ (or, $\mu=\infty$), though this particular choice has no essential consequences for the obtained framework, and other choices of $\mu$ may be easily included. To ensure that we are consistent with this choice throughout the article we must demand that any fiducial vector satisfies,
\begin{align}
\psi_0 (0)=0.
\end{align}
Indeed, one may verify that the eigenvectors of the dilation operator introduced in Eq. (\ref{basis}) satisfy the above condition.

\section*{Appendix B}
We relate the expectation values of the basic observables $\hat{Q}$ and $\hat{P}$ in coherent states $|q,p\rangle$ to the phase space observables $q$ and $p$ by demanding (\ref{physicalcenteringAff}), which in the case of a many-parameter fiducial vector yields two last conditions of Eq. (\ref{physicalcenteringAff3}). Hence, in order for $q$ and $p$ to correspond to the aforementioned expectation values one needs to impose the awkward constraints (\ref{physicalcenteringAff3}) on the fiducial vector labels $\lambda_i\in\mathbb{C}^n$. This problem can be overcome rather easily after one notices that in the absence of the constraints,
\begin{align}
\langle q,p|\hat{Q}|q,p\rangle&=q\langle \psi_0|\hat{Q}|\psi_0\rangle=qQ_{ji}\bar{\lambda}_j\lambda_i,\\
\langle q,p|\hat{P}|q,p\rangle&=p+\langle \psi_0|\hat{P}|\psi_0\rangle=p+P_{ji}\bar{\lambda}_j\lambda_i.
\end{align}
Hence, one may ignore the constraints (\ref{physicalcenteringAff3}) and treat $q$ and $p$ as auxiliary parameters. The genuinely classical observables can be then defined as follows,
\begin{align}
q^s=qQ_{ji}\bar{\lambda}_j\lambda_i,~~
p^s=p+P_{ji}\bar{\lambda}_j\lambda_i.
\end{align}
The quantities $Q_{ji}\bar{\lambda}_j\lambda_i$ and $P_{ji}\bar{\lambda}_j\lambda_i$ are constant along the motion. In the studied examples we therefore first solve the dynamics for $q$ and $p$ and next plot the evolution of $q^s$ and $p^s$. In terms of the geometric viewpoint that we develop at the end of Sec. \ref{restrictedaction}, the employment of fiducial vectors which do not satisfy the constraints (\ref{physicalcenteringAff3}) is equivalent to fixing the respective family of coherent states via a state in a fiber different than $(1,0)$, which is clearly an admissible procedure provided that one recalculates the expectation values as shown above.

\end{document}